\documentclass[prl,twocolumn,superscriptaddress,floatfix,
showpacs,preprintnumbers]{revtex4}

\usepackage[latin2]{inputenc}
\usepackage{amsmath}
\usepackage{amssymb}
\usepackage{epsfig}
\usepackage{bm}
\usepackage{color}

\begin{document}

\title{Coexistence and Transition between Shear Zones in Slow 
Granular Flows}
\author{Robabeh Moosavi}
\affiliation{Department of Physics, Institute for Advanced 
Studies in Basic Sciences, Zanjan 45137-66731, Iran}
\author{M.\ Reza Shaebani}
\affiliation{Computational Physics Group, University of
Duisburg-Essen, D-47048 Duisburg, Germany}
\affiliation{Department of Theoretical Physics, Saarland 
University, D-66041 Saarbr\"ucken, Germany}
\author{Maniya Maleki}
\affiliation{Department of Physics, Institute for Advanced 
Studies in Basic Sciences, Zanjan 45137-66731, Iran}
\author{J\'anos T\"or\"ok}
\affiliation{Department of Theoretical Physics, Budapest 
University of Technology and Economics, Budapest H-1111, 
Hungary}
\author{Dietrich E.\ Wolf}
\affiliation{Computational Physics Group, University of
Duisburg-Essen, D-47048 Duisburg, Germany}
\author{Wolfgang Losert}
\affiliation{Department of Physics, University of Maryland, 
College Park, Maryland, USA}

\date{\today}

\begin{abstract}
We report experiments on slow granular flows in a split-bottom 
Couette cell that show novel strain localization features. 
Nontrivial flow profiles have been observed which are shown to 
be the consequence of simultaneous formation of shear zones in 
the bulk and at the boundaries. The fluctuating band model 
based on a minimization principle can be fitted to the 
experiments over a large variation of morphology and filling 
height with one single fit parameter, the relative friction 
coefficient $\mu_{_\text{rel}}$ between wall and bulk. 
The possibility of multiple shear zone formation is controlled 
by $\mu_{_\text{rel}}$. Moreover, we observe that the symmetry 
of an initial state, with coexisting shear zones at both side 
walls, breaks spontaneously below a threshold value of the shear 
velocity. A dynamical transition between two asymmetric flow 
states happens over a characteristic time scale which depends 
on the shear strength.
\end{abstract}

\maketitle

The intriguing rheology of granular materials has been widely 
studied over the years for its fundamental scientific 
interest and industrial and geophysical importance 
\cite{GranularReview,Schall10,Nedderman92}. Shear banding is 
a widespread phenomenon in slow flows of complex materials, 
ranging from foams \cite{FoamRef} and emulsions 
\cite{EmulsionsRef} to colloids \cite{ColloidsRef} and 
granular matter \cite{Schall10,AvalancheRef,GeoFaultRef,
Mueth00,Bocquet01,Losert00,Latzel03,Fenistein03,Cheng06,
Fenistein06,Unger04,Torok07,Unger07,Borzsonyi09}. A clear 
understanding of how the strain is localized and the material 
yields is crucial in order to develop a consistent continuum 
theory at low inertial numbers, which is currently an 
important open issue \cite{Bocquet09,Bocquet01,Mansard12}. 

Slowly sheared granular materials usually develop narrow shear 
zones, often localized near a boundary, e.g., in avalanches 
\cite{AvalancheRef}, geological faults \cite{GeoFaultRef}, 
and Couette flows \cite{Mueth00,Bocquet01,Losert00,Latzel03}, 
to mention a few. The characteristic length scale of 
the flow gradient is independent of the shear rate, ranges up 
to few particle diameters, and depends on particle shape 
and properties \cite{Mueth00,Bocquet01}. An important question 
we address is whether the formation of boundary-localized 
shear zones is intrinsic to granular matter or whether it 
can be prevented or controlled by suitable boundary conditions. 
Note that wide shear zones in granular bulk flow have been 
created, in a modified split-bottom Couette cell \cite{Fenistein03,
Cheng06,Fenistein06}. The emerging flow profiles were found 
to have shear zones tens of particle diameters wide. The 
wide shear zones were found to obey a number of scaling 
laws, with a transition from a shear zone near the surface 
at low filling heights to a closed cupola shape at high 
filling heights. It has not been clear, so far, whether 
or under what conditions the coexistence of these wide 
shear zones with the boundary-localized ones is possible, 
and what happens to the universality of the flow profiles 
when dealing with more complex boundary conditions. 

In slow flows, i.e.\ the state with rate-independent stresses, 
one expects that the steady-state flow pattern remains stable. 
One of the major findings of the present study is that the 
above concept does not work at shear velocities below a 
critical value.

\begin{figure}[b]
\centering
\includegraphics[scale=0.4,angle=0]{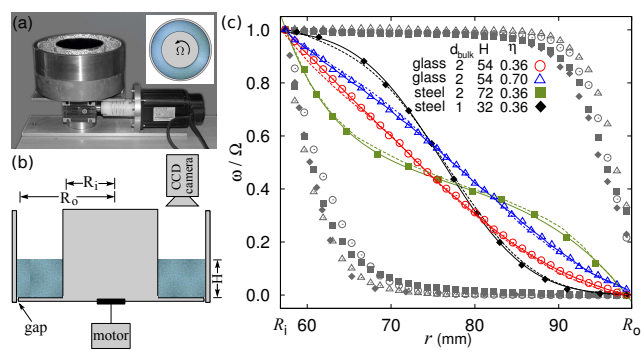}
\caption{(color online). (a) The experimental setup and 
its top view (inset). (b) Schematic of the side view. 
$R_\text{i}{=}57 \text{mm}$, $R_\text{o}{=}99\text{mm}$. 
(c) Velocity profiles at different values of $H$ and 
$d_\text{bulk}$ (in $\text{mm}$), and $\eta$. Symbols 
are experimental data, dashed curves are obtained from 
the variational approach Eq.~(\ref{Eq:1}), and solid 
curves are the fits with Eq.~(\ref{Eq:3}). The gray 
color denotes the localized profiles at the two extreme 
limits of $H$.}
\label{Fig1}
\end{figure}

In this letter we report on the experimental and numerical 
study of complex shear zone formation in a Couette cell 
geometry in which the split at the bottom is located at 
the outer cylinder (see Fig.~\ref{Fig1}). We show that the 
surface flow patterns can be explained by the linear 
combination of three distinct shear zones. Their existence 
is explained by a model based on an optimization principle 
which was already applied to shear zone formation in granular 
materials \cite{Unger04,Torok07,Unger07,Borzsonyi09}. The 
relative magnitude of the bulk and wall effective friction 
coefficients turns out to be the key control parameter 
which determines the possibility of simultaneous formation 
of shear zones in the system and, hence, the overall shape 
of the flow profiles. More interestingly, upon decreasing 
the driving strength below a critical value $\Omega_\text{c}$, 
we observe a dynamical transition between boundary-localized 
shear zones.

{\it Setup ---} The experimental setup is shown in 
Fig.~\ref{Fig1}, with the inner and outer radius, 
$R_\text{i}$ and $R_\text{o}$, respectively. The 
bottom plate and the inner cylinder of the apparatus 
rotate while the outer wall remains at rest. The 
cylindrical gap between the moving and standing parts 
has a size (${<} 400 \,\mu \text{m}$) much smaller than 
the typical grain size, so that no particle can escape. 
The apparatus was filled up to height $H$ with spherical 
glass or steel beads of average diameter $0.5$, $1$, $2$, 
or $3 \,\text{mm}$ with size polydispersity of about $15\%$. 
A layer of grains is glued to the bottom and side walls 
to obtain rough boundaries. The size polydispersity ensures 
that the flow profiles near the walls are not influenced by 
the ordering of grains \cite{Mueth00,Chambon03}. While 
the bulk and boundary beads are always chosen of the 
same material, their size ratio $\delta{=}
d_\text{wall}/d_\text{bulk}$  was varied in order to 
investigate the impact of the relative boundary roughness 
$\eta$ which is defined by the normalized penetration of 
the flowing particles into the rough surface as $\eta {=} 
1 {+} \delta {-} \sqrt{1 {+} 2 \delta {-} \delta^2{/}3}$ 
\cite{Shojaaee12,Koval11}. For smooth walls $\eta {=} 0$.

\begin{figure}[t]
\centering
\includegraphics[scale=0.41,angle=0]{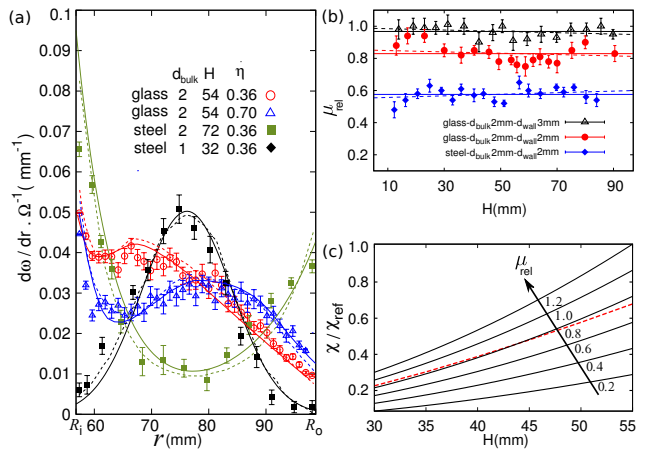}
\caption{(color online). (a) Radial dependence of 
the strain rate. Same symbols and lines as in 
Fig.~\ref{Fig1}(c). (b) $\mu_{_\text{rel}}$ 
obtained from the best fit of Eq.~(\ref{Eq:1}) to 
the data at different $H$. The horizontal solid lines 
indicate the mean values and the dashed lines are 
the best linear fits. (c) The rate of energy dissipation 
$\chi$, scaled by the maximum dissipation of the 
bulk profile $\chi_{_\text{ref}}$, versus $H$. The 
dissipation of the wide shear 
zone (dashed line) is compared to that of the localized 
shear zone at the outer cylinder at different 
values of $\mu_{_\text{rel}}$ (solid lines).}
\label{Fig2}
\end{figure}

{\it Velocity profiles ---}
The inner cylinder and the co-moving bottom plate are 
rotated at angular velocity $\Omega$. To avoid 
rate-dependent stresses \cite{Hartley03}, a gear is used 
to decrease the rotating shaft speed down to the range 
$0.05 \,\text{rad/s} {<} \,\Omega\, {<} 0.15 \,\text{rad/s}$ 
where the steady-state velocities are proportional to 
$\Omega$. Here we show results for $\Omega {=} 0.15 
\,\text{rad/s}$. The resulting surface flow is monitored 
from above using a fast CCD camera with pixel resolution 
$70 \,\mu \text{m}$ at a frame rate of $60 \,\text{s}^{-1}$. 
The average angular velocity $\omega(r)$ at the free 
top surface is obtained by means of particle image 
velocimetry method, which determines the average 
angular cross-correlation function in terms of the 
radial coordinate $r$ for temporally separated frames. 
After the flow reaches a steady state (generally in a 
few seconds), we measure $\omega(r)$ at the free 
surface as a function of $r$. The flow is wall-localized 
for very shallow ($H {\rightarrow} 0$) and deep 
($3(R_o{-}R_i) {<} H$) layers, with exponentially 
decaying strain rates. However, a rich variety of 
surface flow patterns can be observed in the middle 
range of $H$ [see Figs.~\ref{Fig1}(c) and \ref{Fig2}(a)]; 
The profile shapes strongly depend on the choice of $H$, 
$\eta$, and material properties. The basic question is 
how does the system adopt a stationary velocity profile.

{\it Variational approach ---} To provide physical 
insight into what determines the flow profile 
shape, we use a variational minimization procedure 
\cite{VariationRefs}. This method has been successfully 
applied to predict the closed cupola forms of 
shearing regions in deep granular beds 
\cite{Unger04,Cheng06,Fenistein06,Torok07} and 
the refraction of shear zones in layered granular 
materials \cite{Unger07,Borzsonyi09,Knudsen09}. 
Dry granular materials are best described by the 
Mohr-Coulomb theory, which limits the shear stress 
divided by the normal stress by the effective 
friction coefficient $\mu_{_\text{eff}}$ of the 
material. Once the stress ratio exceeds 
$\mu_{_\text{eff}}$, the material fails and a 
shear band forms. Due to cylindrical symmetry the 
whole system can be described by a two dimensional 
radial cut. The resulting shear band must be 
compatible with the boundary conditions and it 
should be the one which fails under the least 
torque or equivalently under the least dissipation 
rate. The last criterion can be formulated as
\begin{eqnarray}
\int_{_0}^{_H} \!\! \mu_{_\text{eff}} \, r(h)^2 (H-h) 
\displaystyle\sqrt{1+(\text{d}r/\text{d}h)^2} \; 
\text{d}h = \text{min},
\label{Eq:1}
\end{eqnarray}
where we search for the $r(h)$ function, i.e.\ 
the shear band position in the bulk of 
material at a given height $h$. Here we used 
hydrostatic pressure since Janssen-effect 
plays no role due to the constant agitation 
of the driving \cite{Nichol10}.

The above plastic event (i.e.\ the instantaneous shear 
band) modifies the structure of the material in its 
vicinity. Hence, due to local fluctuations, another 
shear band can be optimal in the next instance. This 
is thus a self organized process, where the shear 
band appears as a global optimum which itself 
modifies the medium in which the optimization 
is carried out. This is incorporated in a kinetic 
elastoplastic theory \cite{Bocquet09} which takes 
such self organization into account. However, 
it is impossible to solve the model for the 
geometry of our problem, therefore, we 
use a fluctuating band model. The details of this 
model can be found in \cite{Torok07}, here 
we reiterate only the main points: The two 
dimensional cut is coarse grained (coarse graining 
length can be as small as the particle diameter) 
into small mesoscopic cells which are characterized 
by a local effective friction coefficient. The 
friction coefficient is different in the bulk 
$\mu_{_\text{eff}}^\text{bulk}$ and at the wall 
$\mu_{_\text{eff}}^\text{wall}$ due to differences 
in texture. The actual strength of a particular 
cell in the bulk (at the wall) is chosen randomly 
from the interval $[0,\mu_{_\text{eff}}^\text{bulk}]$ 
($[0,\mu_{_\text{eff}}^\text{wall}]$). An instantaneous 
shear band is chosen by minimizing Eq.~(\ref{Eq:1}). 
In the scope of this model, the width of this shear
band is considered to be only one cell wide.
Once the shear band is found, the local strength 
along it and in its neighborhood (next neighbor 
sites) are updated randomly. Shear profiles are 
obtained by an ensemble average over instantaneous 
slips. We note that: (i) The actual probability 
distribution of $\mu_{_\text{eff}}$ \cite{Shaebani08} 
is not important in itself; The central limit theorem 
ensures that only its average and variance play a 
role in the integral of Eq.~(\ref{Eq:1}). 
(ii) The model has other parameters which 
are fixed by the geometry using the coarse graining 
length of a particle diameter size. The only free 
parameter we can vary for a given test is the ratio 
of the friction coefficients of wall and bulk 
$\mu_{_\text{rel}} {=} \mu^\text{wall}_{_\text{eff}} 
/ \mu^\text{bulk}_{_\text{eff}}$. 

The numerical velocity profiles obtained by tuning 
the single free parameter $\mu_{_\text{rel}}$ match 
remarkably with the experimental data, as shown in 
Figs.~\ref{Fig1}(c) and \ref{Fig2}(a), given the fact 
that the boundary roughness is nonuniform, and size 
polidispersity would also influence the mechanical 
properties \cite{Shaebani12}. For a given set of bulk 
and wall particle sizes, the corresponding values of 
$\mu_{_\text{rel}}$ at different filling heights are 
obtained from the best fit to the experimental data 
with Eq.~(\ref{Eq:1}) within $7\%$ error, showing that 
$\mu_{_\text{rel}}$ is roughly invariant with $H$ 
[Fig.~\ref{Fig2}(b)]. The constant nature of 
$\mu_{_\text{rel}}$ indicates that the fluctuating 
band model captures the right physics behind the 
effect of the walls because $\mu_{_\text{rel}}$ is 
fixed by the material size and type on the wall and 
in the bulk, therefore, it should be the same for all 
filling heights for a given set of materials.

\begin{figure}[t]
\centering
\includegraphics[scale=0.38,angle=0]{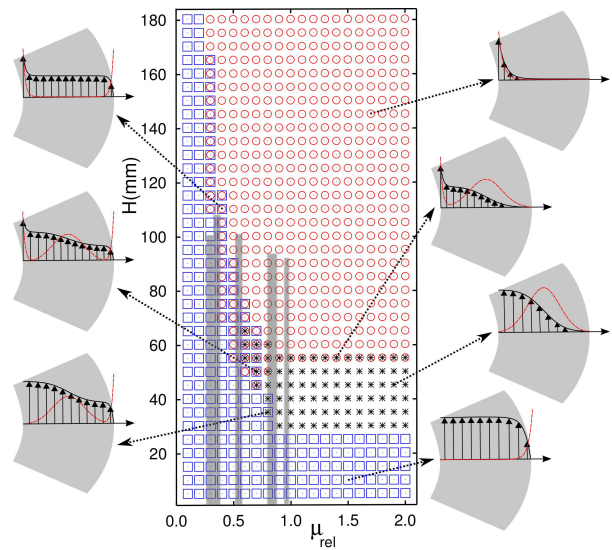}
\caption{(color online). Phase diagram of shear 
zone coexistence for $d_\text{bulk}{=}2\,\text{mm}$. 
The squares, circles, and stars denote the shear 
zone at the outer and inner cylinders and in 
the bulk, respectively. The gray shaded regions 
denote the values of $(\mu_{_\text{rel}}$,$\,H)$ 
for which the experiments are performed. Insets: 
Typical velocity fields (sketched with arrows) 
and the corresponding strain rates (red curves) 
(both corrected for the radial dependence).}
\label{Fig3}
\end{figure}

When looking at different strain rate profiles, up to 
three maxima can be observed, one in the middle and two 
at the boundaries. In the geometry of our setup, these are 
indeed the only feasible choices of shear zones which 
minimize the rate of energy dissipation. The competition 
between these types of minimal paths gives rise 
to a rich shear zone phase diagram. Roughly speaking, 
the energy dissipation along the shear zone at the 
outer cylinder is proportional to 
$\mu^\text{wall}_{_\text{eff}} R^2_\text{o} H^2$, 
while the cost of the path which sticks to the 
bottom plate and then to the inner cylinder grows 
with $\mu^\text{wall}_{_\text{eff}} (R^2_\text{i} H^2{+} 
\frac{2}{3}(R^3_\text{o}{-}R^3_\text{i}) H)$. 
Hence, one expects that the inner shear zone 
wins the race only above $H {\sim} \text{80mm}$.
Assuming that the middle shear zone with the 
center position $R_\text{w}$ is the universal 
wide zone reported in \cite{Fenistein03,Cheng06,
Fenistein06}, it should follow a path in the bulk 
of material which is given by \cite{Unger04} 
\begin{eqnarray}
h = H - r \Big[ 1- \frac{R_\text{o}}{r} 
[1-(H/R_\text{o})^\alpha]\Big]^{1/\alpha},
\label{Eq:2}
\end{eqnarray}
and the total dissipation along the broad shear 
zone is equal to $2\pi \mu^\text{bulk}_{_\text{eff}} 
\int^{^{R_\text{o}}}_{_{R_\text{w}}} (H{-}h) r^2 
\sqrt{1+(dh/dr)^2} \, dr$. The exponent $\alpha$ 
is introduced after Eq.~(\ref{Eq:3}). A comparison 
between this trajectory (within the range of $H$ it 
may exist) and the trajectory which sticks to the 
outer cylinder shows that the former becomes 
favorable only for $\mu_{_\text{rel}} {\gtrsim} 0.8$ 
[Fig.~\ref{Fig2}(c)]. After detailed calculations, 
Fig.~\ref{Fig3} summarizes the results of the 
formation and coexistence of shear zones in a 
phase diagram in the ($\mu_{_\text{rel}}$,$\,H$) 
space. The numerical diagram reveals that the 
$H$-dependence of the surface profile shape has 
a nontrivial dependence on $\mu_{_\text{rel}}$. 
This has been confirmed by the experimental 
results, obtained for the accessible values of 
$(\mu_{_\text{rel}},H)$.

The model numerically  reproduces the experiment well
without providing an explicit analytical expression 
for the velocity profiles. In the following we address 
whether a functional form can be proposed, based on the 
combination of possible basic ingredients: 
wall-localized shear zones with exponential 
flow profiles \cite{Mueth00,Bocquet01,Losert00,Latzel03} 
and wide shear zones with Gaussian velocity gradient 
profiles \cite{Fenistein03,Torok07}. We find that 
both all experimental and numerical profiles are well 
fitted by a superposition of a Gaussian and two 
exponential curves [solid lines in Figs.~\ref{Fig1}(c) 
and \ref{Fig2}(a)]:
\begin{eqnarray}
\frac{\text{d}\omega(r)}{\text{d}r} = a_\text{i} \, \text{exp} 
[-b_\text{i} (r{-}R_\text{i})] + a_\text{o} \, \text{exp} 
[-b_\text{o} (R_\text{o}{-}r)] \nonumber \\
{+} \frac{a_\text{w}}{\sqrt{\pi}\;\xi} \text{exp} 
[-(x-R_\text{w})^2/\xi^2].
\,\,\,\,\,\,\,\,\,\,\,\,\,\,\,\,\,\,\,\,\,\,\,\,\,\,\,
\label{Eq:3}
\end{eqnarray}
The contribution of different terms evolves with $H$ in 
such a way that confirms the validity of the numerical 
phase diagram. Also, the universality of the wide shear 
zone is preserved, i.e.\ the evolution of the width 
$\xi$ and the center position $R_\text{w}$ of the wide 
zone follows, respectively, $H^{2/3}$ and 
$R_\text{o}(1{-}(H/R_\text{o})^{\alpha})$, compatible 
with prior work \cite{Fenistein03,Fenistein06}. 
The exponent $\alpha$, however, ranges between $1.4{-}2.5$. 
The discrepancy can be attributed to the relatively large 
$d_\text{bulk}$ compared to the system size. 

\begin{figure}[t]
\centering
\includegraphics[scale=0.42,angle=0]{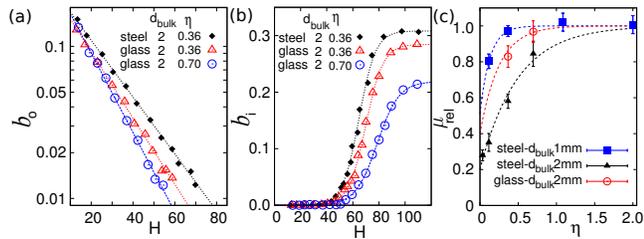}
\caption{(color online). (a),(b) Evolution of the decay 
exponents $b_\text{o}$ and $b_\text{i}$ with $H$. 
The lines indicate exponential fits (a), and fits 
given by Eq.~(\ref{Eq:4}) (b). (c) $\mu_{_\text{rel}}$ 
(averaged over all filling heights) vs.\ the wall 
roughness $\eta$. The curves are exponential 
saturation fits as guides to the eye.}
\label{Fig4}
\end{figure}

We find that our additional parameters, the characteristic 
lengths of the exponential decays $b_\text{i}$ and $b_\text{o}$, 
are influenced by the particle size and type.  They evolve 
with the filling height in the following way: For a given 
experimental setting, $b_\text{o}$ scales with $H$ as 
$\text{exp}({-}\lambda_\text{o}H)$ [Fig.~\ref{Fig4}(a)]. 
The decay constant $\lambda_\text{o}$ grows weakly with 
increasing $\eta$, meaning that the larger roughness is 
accompanied by the faster suppression of the outer shear 
zone with increasing the filling height. The exponent 
$b_\text{i}$ shows a saturation behavior with $H$ 
[Fig.~\ref{Fig4}(b)], with the 
following empirical scaling relation 
\begin{equation}
b_\text{i}(H)/b^{\infty}_\text{i} \sim \Big(1+\text{tanh}
\Big[\frac{H-H_\text{o}}{2\text{w}}\Big]\Big), 
\label{Eq:4}
\end{equation}
with $H_\text{o}$ and $\text{w}$ being the center and 
width of the hyperbolic tangent. The saturation value 
$b^{\infty}_\text{i}$ decays exponentially with $\eta$ 
for a given material (not shown). In short, the surface 
flow pattern is a linear combination of a few basic 
elements, each of which satisfies simple scaling laws.

We also determine the relation between $\mu_{_\text{rel}}$ 
and the boundary roughness $\eta$. As illustrated in 
Fig.~\ref{Fig4}(c), a clear dependency on the material 
type can be observed. One expects that $\mu_{_\text{rel}}$ 
saturates towards $\mu^{\infty}_{_\text{rel}}{=}1$ at 
$\eta {\rightarrow} \infty$, since the bulk particles 
fill the holes and smoothen the boundary roughness so 
that they practically roll over each other. The behavior 
at $\eta {\rightarrow} 0$ depends on material type 
and particle size. We attribute the particle size 
dependence to the roughness caused by the uneven gluing.

\begin{figure}[t]
\centering
\includegraphics[scale=0.35,angle=0]{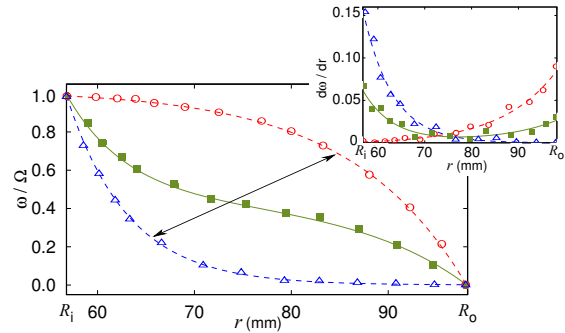}
\caption{(color online). The surface flow profile at 
$\Omega{=} 0.1 \,\text{rad/s}$ (solid curve), and two 
typical profiles at $\Omega{=} 4 {\times} 10^{-3} \, 
\text{rad/s}$ (dashed curves), and the corresponding 
strain rates (inset).}
\label{Fig5}
\end{figure}

{\it Instability at low shear velocities ---} All the 
experimental results reported so far were obtained in 
the rate independent regime, $0.05 \,\text{rad/s} {<} 
\Omega {<} 0.15 \,\text{rad/s}$, where the flow 
profiles rapidly reach to their final steady-state 
shapes. Let us consider a case with two coexisting 
shear zones at both side walls, obtained at $\eta{=}0.36$ 
and by adjusting the height of steel bead layer to $H 
{\simeq}80 \,\text{mm}$. We observe an anomalous 
behavior, a spontaneous symmetry breaking of the 
flow profile, as the shear velocity is decreased 
below $\Omega_{c} {\sim} 5 {\times} 10^{-3} \, 
\text{rad/s}$ (see Fig.~\ref{Fig5}). The system is 
found in either of the two asymmetric flow states 
with strain localization at only one boundary. A 
dynamical transition between the two states takes 
place over a characteristic time scale, which decays 
to zero at $\Omega{\rightarrow}\Omega_{c}$
($\Omega{<}\Omega_{c}$). A similar 
asymmetric shear zone has been recently reported 
in experiments on colloidal glasses \cite{Besseling10} 
(although with permanent rather than transient 
behavior), and in numerical simulations of plane 
shear flow \cite{Shojaaee12}. Based on the analysis 
of velocity fluctuations, a plausible scenario 
is that the agitations induced by the external 
driving at shear velocities lower than $\Omega_c$ are 
not strong enough to trigger shear zones at both walls. 
Thus the system is trapped in one of 
 two minimal states.  The shear rate plays 
the role of a kind of ``temperature", enabling the
system to visit both minimal states. When the system 
is sheared slower than $\Omega_{c}$, it freezes in 
one of the shear zone locations for a long time. 
As the shear velocity approaches $\Omega_{c}$, the 
switching happens more frequently, and the transition 
time goes to zero. Note that the velocity profiles 
in the small $\Omega$ state can be also recovered 
from the fluctuating band model when averaging over 
a long time window. 

In conclusion, the possibility of multiple shear zones 
and the transitions between two of the most thoroughly 
investigated kinds of shear flow behaviors in dry granular 
materials is studied through careful comparison of 
experiment and modeling. We describe those aspects 
of the microstructure that are translated to the global 
rheology, and verify that the formation of localized 
boundary shear zones is not an intrinsic property of 
granular matter. One can adjust the relative strength 
of bulk and boundary shear zones by tuning the relative 
effective friction. Tuning it via  the boundary conditions 
and material properties it is possible to either enhance 
or minimize boundary shear zones.  

Our study may also be used as a template for a 
practical tool to measure the effective friction 
coefficient of the material from surface flow 
patterns. Our finding, that the minimization of 
energy dissipation governs the intriguing behavior, 
is a major step forward towards understanding the 
mechanisms of shear localization in granular 
materials which is an outstanding challenge in 
physics of complex flows, geophysics, and industry. 
The observed instabilities at low shear velocities 
deserves further detailed studies to uncover the 
underlying physics. The results of plane shear 
flows \cite{Shojaaee12} suggest that the rotational 
degrees of freedom of particles play a crucial 
role in facilitating the dynamical transitions 
between the optimum states.

We would like to thank T. Unger and Z. Shojaaee 
for helpful discussions. R.M. and M.M. acknowledge 
the support of this project by the Institute for 
Advanced Studies in Basic Sciences (IASBS) Research 
Council under grant No. G2010IASBS136, and M.R.S. 
and D.E.W. by DGF Grant No.\ Wo577/8-1 within the 
priority program ``Particles in Contact''. WL 
acknowledges support from DTRA grant 10DTRA1077.

\end{document}